\documentclass[a4paper,fleqn,usenatbib]{mnras}
\usepackage{newtxtext}
\usepackage{soul}
\usepackage[T1]{fontenc}
\usepackage{ae,aecompl}
\usepackage{graphicx}	
\usepackage{amsmath}	
\usepackage{amssymb}	
\usepackage{rotating}
\usepackage{lscape}
\usepackage{array}
\usepackage{wrapfig}
\usepackage{graphics}
\usepackage{hyperref}
\usepackage{color}
\usepackage{wasysym}

\usepackage{subfig}

\def\hi{H\,{\sc i}}

\def\kms{km~s$^{-1}$}
\def\msun{M$_{\astrosun}$}

\title[MeerKAT observations of NGC~1512/1510]{MeerKAT HI-line observations of the nearby interacting galaxy pair NGC~1512/1510}

\author[E. C. Elson et al.]{
E. Elson$^{1}$\thanks{E-mail: elson.e.c@gmail.com (ECE)},
M. Glowacki$^{1,2,3}$,
R. Deane$^{4}$,
N. Isaacs$^1$,
X. Ndaliso$^3$
\\
$^{1}$Department of Physics $\&$ Astronomy, University of the Western Cape, Robert Sobukwe Rd, Bellville, 7535, South Africa.\\
$^{2}$Inter-University Institute for Data Intensive Astronomy, Bellville 7535, South Africa\\
$^{3}$International Centre for Radio Astronomy Research, Curtin University, Bentley, WA 6102, Australia\\
$^{4}$Wits Centre for Astrophysics, School of Physics, University of the Witwatersrand, 1 Jan Smuts Avenue, Johannesburg, South Africa\\
$^5$Department of Physics, University of Pretoria, Hatfield, Pretoria, 0028, South Africa}

\begin{document}
\label{firstpage}
\pagerange{\pageref{firstpage}--\pageref{lastpage}}
\maketitle

\begin{abstract}
{We present MeerKAT \hi\ line observations of the nearby interacting galaxy pair NGC~1512/1510.  The MeerKAT data yield high-fidelity image sets characterised by an excellent combination of high angular resolution ($\sim 20''$) and and sensitivity ($\sim 0.08$~\msun~pc$^{-2}$), thereby offering the most detailed view of this well-studied system's neutral atomic hydrogen content, especially the \hi\ co-located with the optical components of the galaxies.  The stellar bulge and bar of NGC~1512 are located within a central \hi\ depression where surface densities fall below 1~\msun~pc$^{-2}$, while the galaxy's starburst ring coincides with a well-defined \hi\ annulus delimited by a surface density of 3~\msun~pc$^{-2}$.  In stark contrast, the star-bursting companion, NGC~1510, has its young stellar population precisely matched to the highest \hi\ over-densities we measure ($\sim12.5$~\msun~pc$^{-2}$).  The improved quality of the MeerKAT data warrants the first detailed measurements of the lengths and masses of the system's tidally-induced \hi\ arms.  We measure the longest of the two prominent \hi\ arms to extend over $\sim 275$~kpc and to contain more than 30$\%$ of the system's total \hi\ mass.   We quantitatively explore the spatial correlation between  \hi\ and far-ultraviolet flux over a large range of \hi\ mass surface densities spanning the outer disk.  The results indicate the system's \hi\ content to play an important role in setting the pre-conditions required for wide-spread, high-mass star formation.  This work serves as a demonstration of the remarkable efficiency and accuracy with which MeerKAT can image nearby systems in \hi\ line emission.}  
\end{abstract}


\section{Introduction}
Interactions between galaxies serve as one of the strongest drivers of their evolution \citep{toomre_1972, gehrz_1983, farouki_1982, moore_1996, hopkins_2008}.  In the hierarchical formation scenario, interactions are ubiquitous and are therefore important to study in order to understand this crucial aspect of the evolutionary process.  Neutral atomic hydrogen (\hi) is a particularly useful tracer of galaxy interactions.  Galaxies typically have \hi\ disks that are a factor $\gtrsim 2$ larger than their stellar disks (\citealt{swaters_2002, toribio_2011, wang_2016}), making \hi\ more susceptible to long-range tidal forces.  Furthermore, given the long rotation periods of the outer disk, the \hi\ distribution and kinematics serve as a long-lasting record of the interaction history.  During interactions, significant fractions of \hi\ can be removed from regions centred on the optical disks of galaxies, and transported to much larger radii.  In the local Universe there are many examples of multiple systems with disturbed gas morphologies consisting of long bridges and tails  (e.g., \citealt{kregel_sancisi_2001, namumba_2021}), as well as being accompanied by gas-rich companions such as dwarf galaxies and/or \hi\ cloud complexes \citep{sancisi_2008}.  These all serve as evidence of recent or ongoing interaction processes.  



The enhanced imaging capabilities of new radio telescopes warrants the study of \hi\ in galaxies spanning a range of intrinsic properties and environments.  In this work, we utilise the unique imaging capabilities of the MeerKAT telescope to map the \hi\ content of the nearby interacting galaxy pair NGC1512/1510.  The system has been extensively studied at several wavelengths and is well-known to have a very extended \hi\ distribution dominated by two tidally-induced \hi\ arms.  Given the system's southern declination, it is ideally-suited to be observed with MeerKAT.  In this work, we produce new \hi\ maps that benefit from an unprecedented combination of high angular resolution and surface brightness sensitivity, and use them to improve our understanding of the distribution of \hi\ in NGC~1512/1510.

Optically, NGC~1512 is classified as type SB(a)a by \citet{de_vaucouleurs_1976}.   It consists of a prominent bulge and bar enclosed in a starburst ring of major axis  $\sim 16$~arcsec.  Deep optical imaging carried out by \citet{hawarden_1979} revealed the presence of extended arms and filaments in NGC~1512, and was taken to indicate the tidal effects of an ongoing gravitational interaction with NGC~1510 - a blue compact dwarf at a projected angular distance of $\sim 5$~arcmin from NGC~1512. Single-dish \hi\ observations of NGC1512/1510 were made by \citet{hawarden_1979} using the Parkes radio telescope.  Their maps showed the \hi\ emission to be centred on NGC~1512 and to extend to a radius of about 60~kpc.  The first \hi\ interferometric observations of the system we carried out by \citet{koribalski_2009} using the Australia Telescope Compact Array (ATCA).  Their images offered the first spatially-resolved views of some parts of the system's extended \hi\ distribution.  In this work, we use our new data to generate a refined view of the spatial distribution of \hi\ in NGC~1512/1510.

Ultra-violet observations of NGC~1512/1510 \citep{gil_de_paz_2007a} also offer clear evidence of a dramatic interaction history.  NGC~1512 has a structured, UV-bright disk extending far beyond its optical disk \citep{thilker_2007}.  The UV disk highlights the high level of recent, high-mass star formation activity occurring due to the tidal interaction between the two galaxies.  For $\gtrsim 200$ stellar clusters in the extended disk, \citet{koribalski_2009} showed \hi\  to be a good tracer of star formation activity, and stated that regions of higher \hi\ surface density have higher star formation rates.  Their result demonstrates how, even for interacting systems such as NGC~1512/1510, star formation can be linked to \hi\ content over a large range of spatial scales.  Similar conclusions have been drawn by other authors (e.g., \citealt{bigiel_2010}) for other nearby galaxies.  \citet{bacchini_2019} presented empirical star formation laws of disk galaxies based on measurements of the volume densities of their gas and star formation rates. They announced the surprising discovery of an unexpected correlation between the volume densities of \hi\ and star formation rate.  In this work, we use our new high-resolution \hi\ images of NGC~1512/1510 to carry out a quantitative study of the coincidence between \hi\ and recent, high-mass star formation.

The layout of this paper is as follows.  In Section~\ref{sec_data} we present the details of the data acquisition and reduction processes.  We discuss our assumed distance measure for the NGC~1512/1510 system in Section~\ref{sec_distance}.  Our various \hi\ data products are presented and discussed in Section~\ref{sec_HIproducts}.  In this section we also carry out a detailed study of the distribution of \hi\ in NGC~1512/1510.  A study of the links between \hi\ content and star formation is given in Section~\ref{SF_laws}.  Finally, a summary of this work is offered in Section~\ref{sec_summary}.

\section {Data reduction and processing}\label{sec_data}
The NGC~1512/1510 system was observed on 18 and 24 May 2019 as part of the 2019 call for `open time'  observing proposals on the MeerKAT radio telescope.  The reader is referred to \citet{jonas_2016}, \citet{camilo_2018} and \citet{mauch_2020} for detailed information regarding the telescope.  Of the 64 MeerKAT dishes, 58/60 were used on 18/24 May, with each day yielding 5 hours 58 minutes of on-target observations.   The data were taken using MeerKAT's L-band receiver which spans the frequency range 900 -- 1670~MHz.  The full bandwidth was split into 4096 channels each of width 208.98~kHz.  At $z=0$ the corresponding velocity width of a channel is $dv\approx 44$~\kms.  At 1420~MHz the MeerKAT field of view is approximately a degree in diameter.  Thus, a single pointing centred on the optical position of NGC~1512 was sufficient to observe the entire NGC~1512/1510 system.  J0440-4333 was used as a time-varying complex gain calibrator and was observed for slightly less than 3 minutes every 15 minutes.  J0408-6545 was used as an absolute flux density and bandpass calibrator and was observed for approximately 10 minutes every 2.5 hours. Table~\ref{data_table} summarises the main aspects of the MeerKAT data used in this study. 

\begin{table}
	\centering
	\caption{Details of MeerKAT observations.}
	\label{data_table}
	\begin{tabular}{ccc} 
		\hline
		parameter						& 	value			&	units\\
		\hline
		Target							&	NGC~1512/1510	&	\\
		Observation dates					&	18, 24 May 2019	&	\\	
		Total observation time				&	15.8				&	h\\
		Total on-target time					&	12				&	h\\
		Frequency range					&	$900-1670$		&	MHz\\
		Central frequency					&	1415				& 	MHz\\
		Number of channels					&	4096				&	\\
		Frequency resolution					&	208.98			&	kHz\\
		Velocity resolution					&	44.1				&	\kms\\
		Bandpass/flux density calibrator		&	J0408-6545		&	\\
		Gain calibrator						&	J0440-4333		&	\\
		rms in a channel 					&	0.11				&	mJy/bm\\
		Angular resolution 					& 	$26.8\times 20.1$	&	$''\times''$\\
		Spatial resolution 					& 	$1.52\times 1.14$	&	kpc$\times$kpc\\
		Pixel scale 						& 	7.5 				&	$''$\\
		\hline
	\end{tabular}
\end{table}

\begin{figure}
\centering
	\includegraphics[width=\columnwidth]{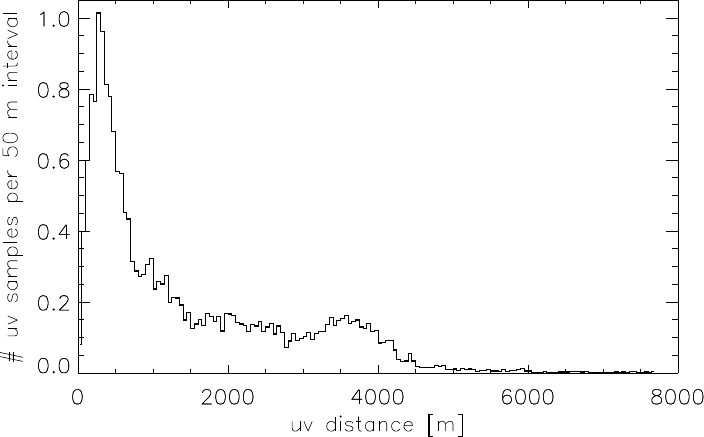}
    \caption{Number of $uv$ samples per 50 m baseline interval for the full 12 hours of on-target time.  The number of samples per bin is given in units of $10^8$.  Baselines shorter than 1~km constitute 52.3~per~cent of the $uv$ coverage.  The maximum baseline length is $\sim 7.6$~km.}
    \label{uv_coverage}
\end{figure}

From the full data set, a frequency window of approximately 30~MHz (150 channels) centred on 1415~MHz was split off and used for the purposes of this study.  Calibration of the MeerKAT data was carried out using the processMeerKAT software developed and maintained at the Inter-University Institute for Data Intensive Astronomy.  The software uses CASA \citep{CASA} tasks and helper functions to carry out the calibration.  Standard cross-calibration, including delay calibration, bandpass calibration, and gain calibration, was applied to the 30~MHz data set.  The excellent $uv$ coverage offered by MeerKAT is shown in  Figure~\ref{uv_coverage}.  Baselines shorter than 1~km constitute 52.3~per~cent of the $uv$ data.  The maximum baseline length is $\sim 7.6$~km.  Throughout this work, The Cube Analysis and Rendering Tool for Astronomy (CARTA, \citealt{CARTA}) was used to inspect the data in various ways.

The CASA task {\sc uvcontsub} was used to subtract a 2nd order polynomial from the line-free channels of the $uv$ data.  The  {\sc tclean} task was used with a Briggs weighting scheme to image each channel of the continuum-subtracted cube.   We first produced a dirty cube by imaging with zero {\sc clean} iterations.   The rms of the noise in a line-free channel of the dirty cube was measured and then used to generate a mask cube in which all emission below/above $3\times \mathrm{rms}$ was set to 0/1.  A final cube was then generated by re-imaging the $uv$ data and using the mask cube to {\sc clean} down to a level of $0.3\times \mathrm{rms}$.  The {\sc clean} task also restored the {\sc clean}ed cube with a Gaussian approximation of the central lobe of the point spread function.  Finally, we applied a primary beam correction to the cube.   We experimented with a range of weighting parameters when producing our data cube.  For this study, we selected the cube based on a robust value $R=1.5$ as being the one that offers a good balance between dynamic range and spatial resolution. The cube has a spatial resolution $26.8''\times 20.1''$ and a pixel scale of  $7.5''$.  The rms of the flux in a line-free channel is 0.11~mJy~beam$^{-1}$ (corresponding to an \hi\ mass surface density of $\sim 0.08$~\msun~pc$^{-2}$).

\section {Distance estimate}\label{sec_distance}
We require an accurate distance measure for the NGC~1512/1510 system in order to convert observed fluxes to luminosities.  The system was observed as part of the Legacy ExtraGalactic UV Survey (LEGUS, PI Calzetti, GO-13364).  \citet{sabbi_2018}  derived the distance from the tip of the red giant branch (TRGB) for each LEGUS galaxy.  \citet{sabbi_2018} warn that due to severe star crowding in the central pointing of NGC~1512  it was not possible to determine the luminosity of the TRGB from the data.  Therefore, they assumed the average values estimated from their south west pointing of NGC~1512 and NGC~1510.  The TRGB distance estimate for NGC~1512 presented by \citet{sabbi_2018} is $D=11.7\pm1.1$~Mpc.   

As part of the HIPASS Bright Galaxy Catalogue, \citet{HIPASS_BGC} present a systemic velocity measure of $V_\mathrm{sys}~=~898$~\kms\ for the NGC~1512/1510 system based on its integrated HI spectrum.  Converting this velocity measure directly to a distance using Hubble's Law  with $H_\mathrm{0}=74.03$~\kms~Mpc$^{-1}$ \citep{riess_2019} yields a Hubble flow distance of 12.13~Mpc, which is within the error bounds of the TRGB distance estimate from \citet{sabbi_2018}.  However, closer agreement is achieved if the systemic velocity and equatorial coordinates of NGC~1512 are entered into the Cosmicflows-3 Distance-Velocity Calculator \citep{CF3_calculator} which takes into account relationships between the distances and velocities of galaxies based on smoothed versions of the velocity fields derived by the Cosmicflows program.  This calculator yields a distance measure of 11.84~Mpc, which is very close to the TRGB estimate of $11.7\pm1.1$~Mpc from \citet{sabbi_2018}.  Hence, in this work, we adopt the \citet{sabbi_2018} distance measure for the entire NGC~1512/1510 system.  We propagate the 1.1~Mpc uncertainty estimate to all quantities based on it. 

\section {Results}\label{sec_HIproducts}
The NGC~1512/1510 system has a remarkable \hi\ morphology.  Our MeerKAT observations benefit from excellent $uv$ coverage, yielding \hi\ data products that are of high spatial resolution and which are also highly sensitive over a large dynamic range to structures spanning a range of angular scales.  They provide us with a new detailed view of the \hi\ in this interacting galaxy pair. In the sections that follow, we present and discuss several detailed measurements of the \hi\ content of the system.  

\subsection{Channel maps}
\begin{figure*}
\centering
	\includegraphics[width=2.2\columnwidth]{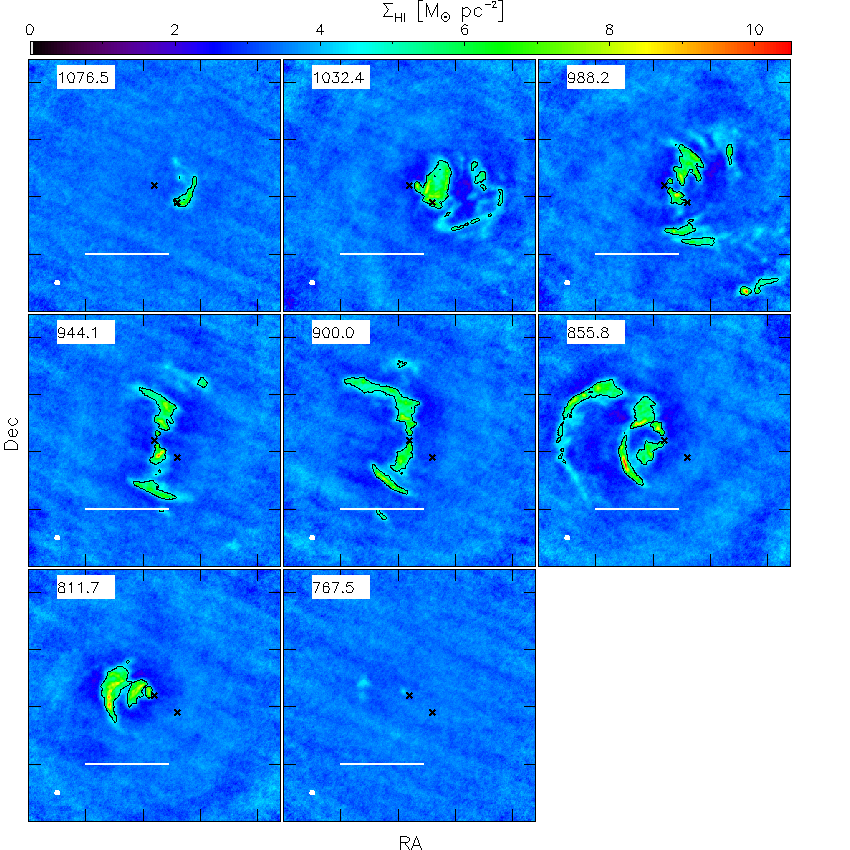}
    \caption{Channel maps of the MeerKAT \hi\ data cube.  In order to present the full dynamic range of the cube, we have boosted the cube by an additive constant of 1.12~\msun~pc$^{-2}$, thereby making all pixels positive (with a minimum mass surface density equal to 0~\msun~pc$^{-2}$).  Then, we have applied a square-root stretch to the boosted mass surface densities.  The colour bar at the top therefore represents \hi\ mass surface density in units of \msun~pc$^{-2}$, stretched in the above-mentioned manner.  Our total intensity map shown in Fig.~\ref{mom0_panel} shows the flux distribution using a linear stretch.  The black contour in each panel is at a level of 1~\msun~pc$^{-2}$ in the original cube.  The rms \hi\ mass surface density of a line-free portion of the original cube is 0.08~\msun~pc$^{-2}$.  The line-of-sight velocity corresponding to each channel is shown in the top left of each panel.  The optical positions are of NGC~1512 and NGC~1510 are represented by the upper and lower black crosses, respectively.  Shown as a grey-filled ellipse in the bottom left of each panel is the $26.8''\times 20.1''$ restoring beam of the cube.  The solid white bar in each panel represents an angular length of  approximately 15 arcmin which corresponds to a physical length of 50 kpc for $D=11.7$~Mpc.  Tick marks along each axis are separated by 10~arcmin.}
    \label{chan_maps}
\end{figure*}

Channel maps for our \hi\ data cube are shown in Figure~\ref{chan_maps}.  In order to present the full dynamic range of the cube, we have boosted it by an additive constant of 1.12~\msun~pc$^{-2}$, thereby making all pixels positive (with a minimum mass surface density equal to 0~\msun~pc$^{-2}$).  Then, in order to highlight fainter emission, we have applied a square root colour stretch in  Figure~\ref{chan_maps} to the boosted mass surface densities.  The colour bar at the top of Figure~\ref{chan_maps} therefore represents \hi\ mass surface density in units of \msun~pc$^{-2}$, colour stretched in the above-mentioned manner.  Each channel spans a velocity range $dv\approx 44$~\kms.  The full velocity range spanned by the system's emission is $\sim 310$~\kms.  Despite the interaction between NGC~1512 and NGC~1510, the overall \hi\ kinematics of the  system are dominated by rotation. 

\subsection{Global profile}
\begin{figure}
\centering
	\includegraphics[width=1\columnwidth]{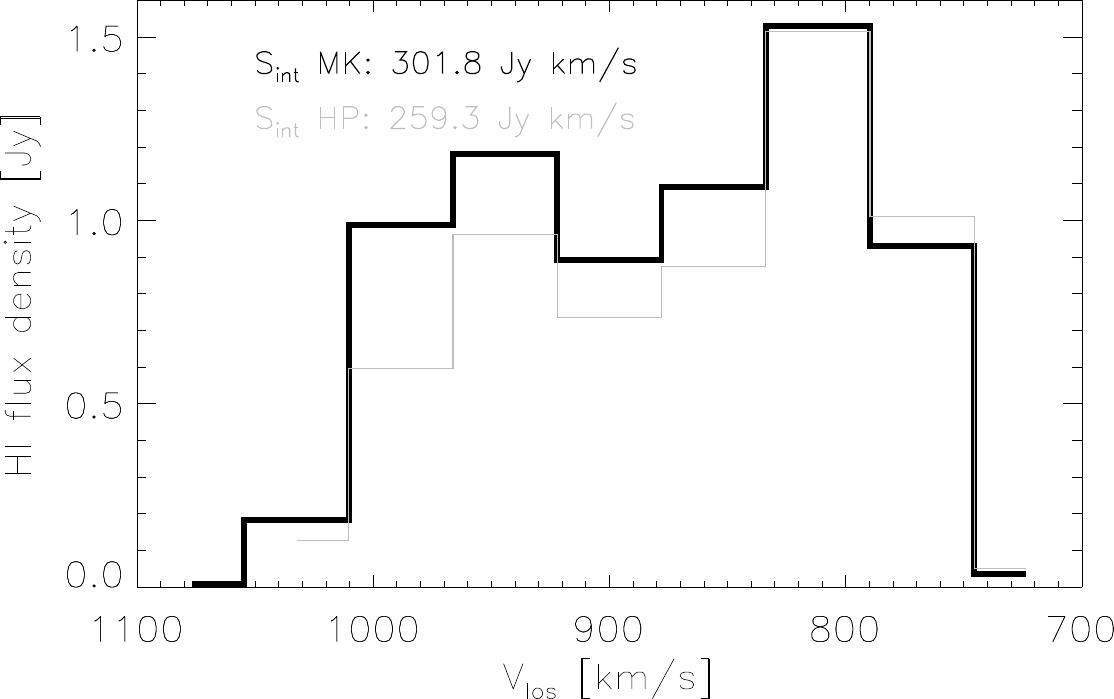}
    \caption{Spatially-integrated \hi\ flux density as a function of line-of-sight velocity for the NGC~1512/1510 system based on our MeerKAT imaging (black) and HIPASS imaging (grey).  The HIPASS spectrum has been re-sampled from its native 13.2~\kms\ velocity resolution to the 44~\kms\ channel size of our MeerKAT data cube.  Shown in the top left of the panel is the total flux density for the galaxy based on the two data sets.  }
    \label{HI_spectrum}
\end{figure}

Figure~\ref{HI_spectrum} shows as a black histogram the spatially-integrated \hi\ profile of the NGC~1512/1510 system.  Despite the coarse velocity resolution, a double-horn profile is clearly visible.  Our cube yields an integrated \hi\ flux density $S_\mathrm{int}=301.8$~Jy~\kms.  For a distance $D=11.7\pm 1.1$~Mpc, this corresponds to a total \hi\ mass $M_\mathrm{HI}=9.74^{11.7}_{8.01}~\times~10^9$~\msun.  Also shown in Figure~\ref{HI_spectrum}, as a grey histogram, is the HIPASS integrated \hi\ spectrum.  To facilitate a direct comparison to our data, the HIPASS spectrum has been re-sampled from its native 13.2~\kms\ velocity resolution to the 44~\kms\ channels of our MeerKAT data cube.  The integrated \hi\ flux density from the HIPASS data is $S_\mathrm{int}=259.3\pm 17.4$~Jy~\kms\ \citep{HIPASS_BGC}.  The ATCA imaging from \citet{koribalski_2009} (based on a total of $\sim 75$ hours of on-source time) yields $S_\mathrm{int}=268$~Jy~\kms.



\subsection{\hi\ total intensity map}\label{mom0_maps}
\begin{figure*}
\centering
	\includegraphics[width=1.75\columnwidth]{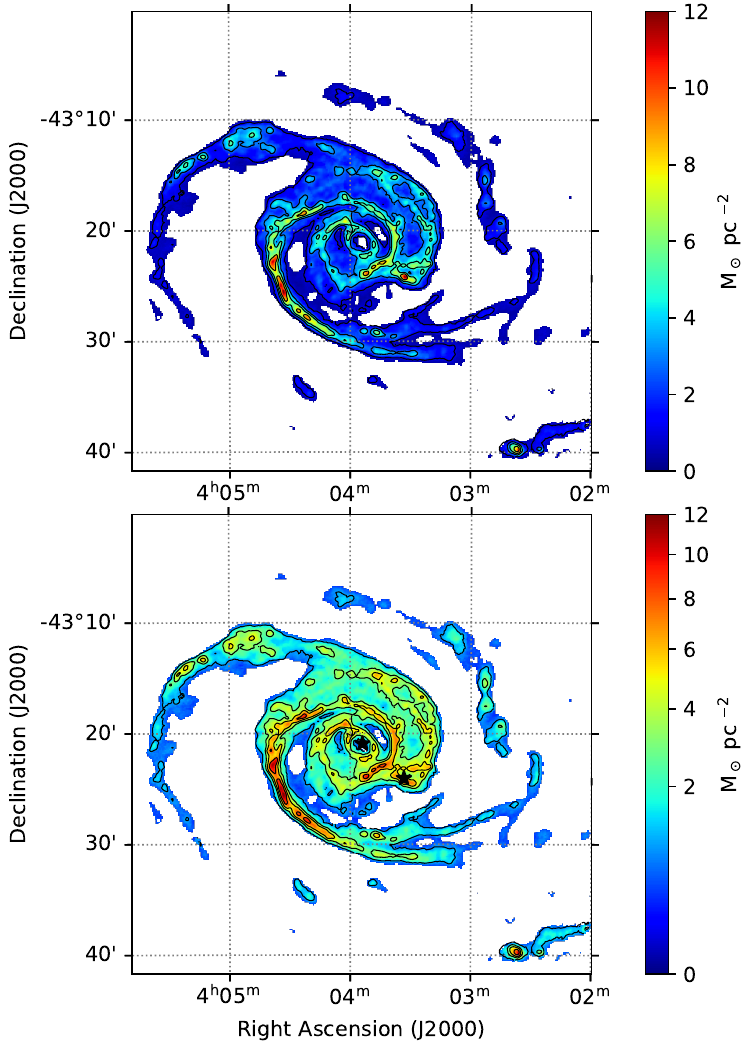}
    \caption{\hi\ total intensity map generated from our MeerKAT data cube based on a linear colour stretch (top panel) and a square-root colour stretch (bottom panel).  For each panel, the colour bar represents the \hi\ mass surface density in units of \msun~pc$^{-2}$.  Overlaid black contours are at levels of 1, 3, 5, 8~\msun~pc$^{-2}$.  The full range of \hi\ mass surface densities is $\sim 0.5$~-~12.5~\msun~pc$^{-2}$.  The optical positions of NGC~1512 and NGC~1510 are indicated by the black stars in the bottom panel.  For each panel, each axis spans a physical length of $\sim 141$~kpc assuming a distance of 11.7~Mpc.} 
    \label{mom0_panel}
\end{figure*}

Our primary data product in this paper is a new \hi\ total intensity map for NGC~1512/1510, possibly serving as the best such map ever produced for the system.  We generated the map by spatially smoothing the \hi\ data cube to a resolution twice that of its native resolution (i.e., to $52''\times 40''$), applying a $3\sigma$ cut in order to remove the Gaussian-distributed noise, and then using the surviving voxels of the smoothed cube to produce a mask that was then applied to the full-resolution cube, which was then spectrally integrated.  Figure~\ref{mom0_panel} presents two views of the \hi\ total intensity map.  The top image uses a linear colour stretch of the \hi\ surface densities to highlight the dense ridges of the extended spiral arms, while the bottom image uses a square-root  stretch to show more detail at lower mass surface densities.  

Immediately striking is the variety of structures present in our map.  It reveals a highly complex \hi\ morphology spanning surface densities from $\sim 0.25$~\msun~pc$^{-2}$ to  $\sim 12.5$~\msun~pc$^{-2}$, making for a dynamic range of approximately 50.  High-intensity \hi\ emission is restricted mainly to the system's two prominent spiral arms which consist of well-defined central \hi\ ridges with surface densities $\gtrsim 3$~\msun~pc$^{-2}$, as well as compact clumps with \hi\ surface densities $\gtrsim 8$~\msun~pc$^{-2}$ spread over regions extending several kiloparsecs.  The disk-like inner  \hi\ distribution  centred on NGC~1512 is dominated by the system's two prominent spiral arms which can now be traced inwards all the way to the starbust ring of NGC~1512.  Our map shows the eastern \hi\ arm to be significantly longer (in a southern direction) than seen in the ATCA images presented in \citet{koribalski_2009}.  The other \hi\ arm that extends south and then wraps around to the west is clearly seen to bifurcate at a declination of approximately $-43^{\circ} 30'$.  After this bifurcation, there is a thin \hi\ bridge extending to the north-west, beyond which the arm then wraps through another $\sim 90^{\circ}$ all the way to a point directly north of the optical disk of NGC~1512.   

 Our map highlights the established fact that the system's \hi\ content is distributed over a large area.  The angular separation between the optical centres of the two galaxies is approximately 5~arcmin, which corresponds to a physical separation of $\sim17$~kpc assuming $D=11.7$~Mpc.  The \hi\ extent of the system, however, is larger than 100~kpc.  Using the \citet{wang_2016} \hi\ mass-size relation to convert our measured total \hi\ mass of $M_\mathrm{HI}=9.74\times 10^9$~\msun\ to an \hi\ diameter yields $D_\mathrm{HI}\sim 56.9$~kpc.  Hence, the observed distribution of \hi\ is approximately twice as large as what is expected for its \hi\ mass.  
 
\begin{figure*}
\centering
	\includegraphics[width=2\columnwidth]{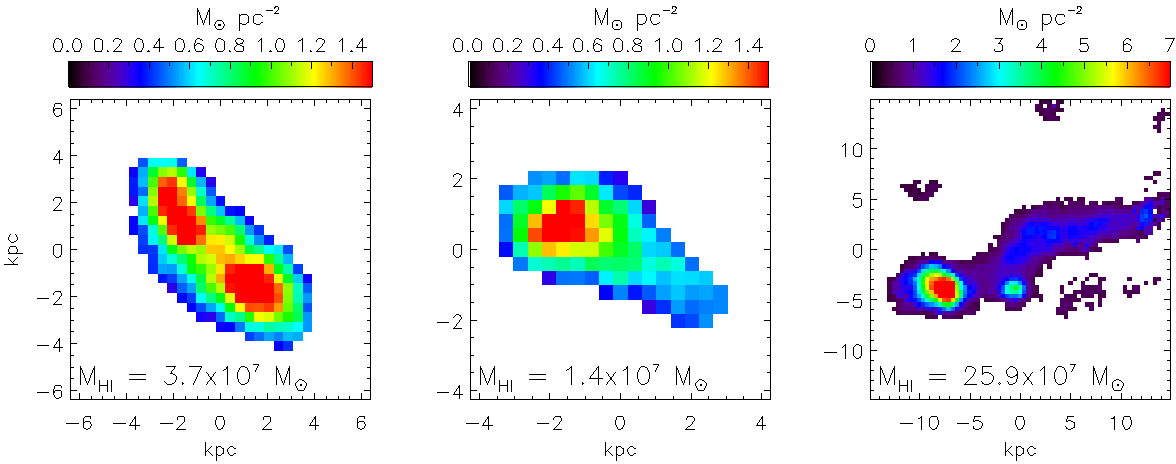}
    \caption{\hi\ clouds seen in the southern part of our  \hi\ total intensity map.  Note the different range of \hi\ mass surface densities (represented in units of \msun~pc$^{-2}$ by the colour bar) shown in each panel.  The \hi\ mass of each is shown at the bottom of its panel.}
    \label{HI_clouds}
\end{figure*}

Three \hi\ clouds are located in the southern part of our total intensity map.  Zoomed-in views of these clouds are shown in Figure~\ref{HI_clouds}.  The shape of each cloud is elongated in a direction that is consistent with the overall spiral structure of the NGC~1512/1510 system.  This suggests the existence of the clouds to be as a direct result of the system's interaction history.  The first two have similar surface density distributions, peaking at values of $\sim 1.5$~\msun~pc$^{-2}$.  The cloud in the south-west corner is likely a tidal dwarf galaxy in the process of forming.  Its mass of $2.59\times 10^8$~\msun\ is close to 3~per~cent of the system's \hi\ mass,  and its mass surface densities peak at $\sim 7$~\msun~pc$^{-2}$.   Its high-density core is separated from the optical centre of NGC~1512 by approximately 79~kpc.  The appearance of the cloud's emission in the channel maps of our data cube shows its kinematics to be consistent with the overall circular rotation of the total \hi\ component.  

\begin{figure}
\centering
	\includegraphics[width=1.\columnwidth]{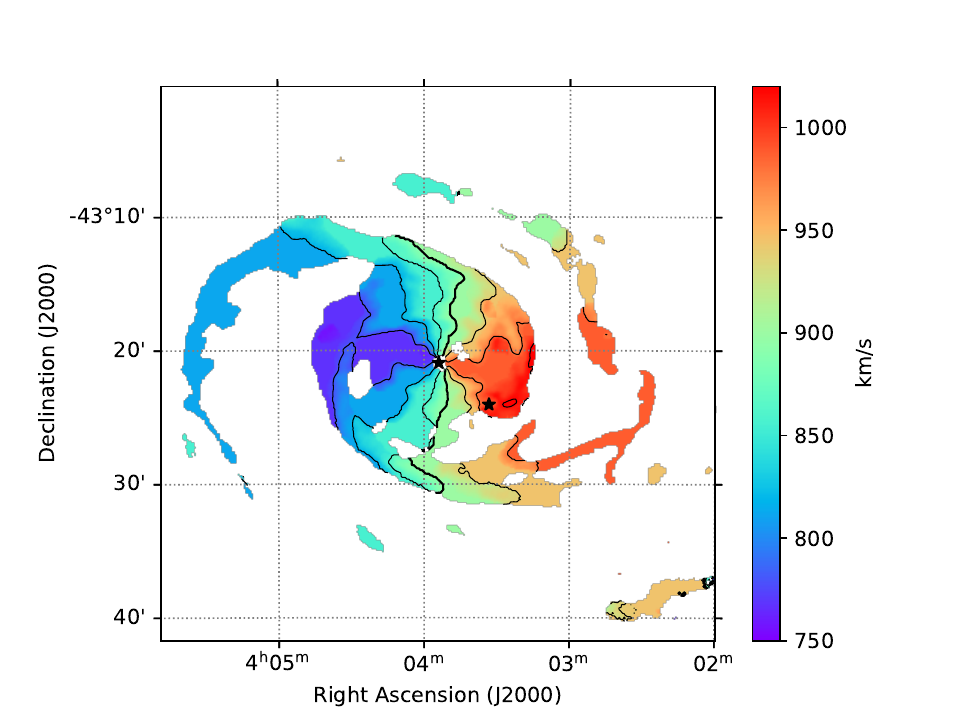}
    \caption{Intensity-weighted mean \hi\ velocity field of NGC~1512/1510. The colour bar represents the line-of-sight velocity in units of \kms.  The thick contour marks the systemic velocity of 898~\kms\ stated by \citet{koribalski_2009}.  Contours are spaced by 40~\kms.  The optical positions of NGC~1512 and NGC~1510 are indicated by the black stars.} 
    \label{mom1}
\end{figure}

For completeness, we show the intensity-weighted mean \hi\ velocity field of the system in Figure~\ref{mom1}.  In spite of the coarse velocity resolution of our \hi\ data cube, the velocity field  exhibits clear signs of regular rotation. We do not attempt to fit a dynamical model to the velocity field.  

\subsection{Expected \hi\ content}
We can contextualise our measurement of the total \hi\ mass for NGC~1512/1510 by comparing it to the sum of the \hi\ masses we expect each galaxy to have.  A typical galaxy has its \hi\ mass correlated to its stellar mass.  \citet{denes_2014} determine scaling relations between the logarithm of the observed \hi\ masses of galaxies and their magnitudes in six different wavebands.  For the $B$-band, the relation is 
\begin{equation}
\log M_\mathrm{HI}=(2.89\pm0.11) -(0.34\pm0.01) M_\mathrm{B}.
\label{denes_eqn}
\end{equation}

From the \citet{gil_de_paz_2007a} data presented in \citet{koribalski_2009} we know the $B$-band apparent magnitudes of NGC~1512 and NGC~1510 to be $11.08\pm 0.09$~mag and $13.47 \pm 0.11$~mag, respectively.  Our assumed distance $D~=~11.7\pm 1.1$~Mpc then yields respective absolute magnitudes of $-19.26^{-19.54}_{-18.95}$~mag and $-16.87^{-17.17}_{-16.54}$~mag.  Inserting these magnitudes into Equation~\ref{denes_eqn} yields $9.43^{9.84}_{9.03}$ and $8.62^{9.01}_{8.24}$ as the logarithms of the \hi\ masses (in units of \msun) of NGC~1512 and NGC~1510, respectively.   Hence, the expected combined \hi\ mass for the two galaxies is $3.16^{7.96}_{1.25}\times 10^9$~\msun.  This estimate is a factor $\sim 3$ lower than  our measurement of $M_\mathrm{HI}~=~9.74^{11.7}_{8.01}~\times~10^9$~\msun\ for the total \hi\ mass.  \citet{denes_2014} consider a galaxy to be anomalous in terms of its \hi\ content only if it differs by more than a factor $\sim 4$ from the mean relation.  Nevertheless, it seems clear that NGC~1512/1510 has an excess of \hi\ mass. 

\subsection{Inner \hi\ distribution}
\begin{figure*}
\centering
	\includegraphics[width=1.8\columnwidth]{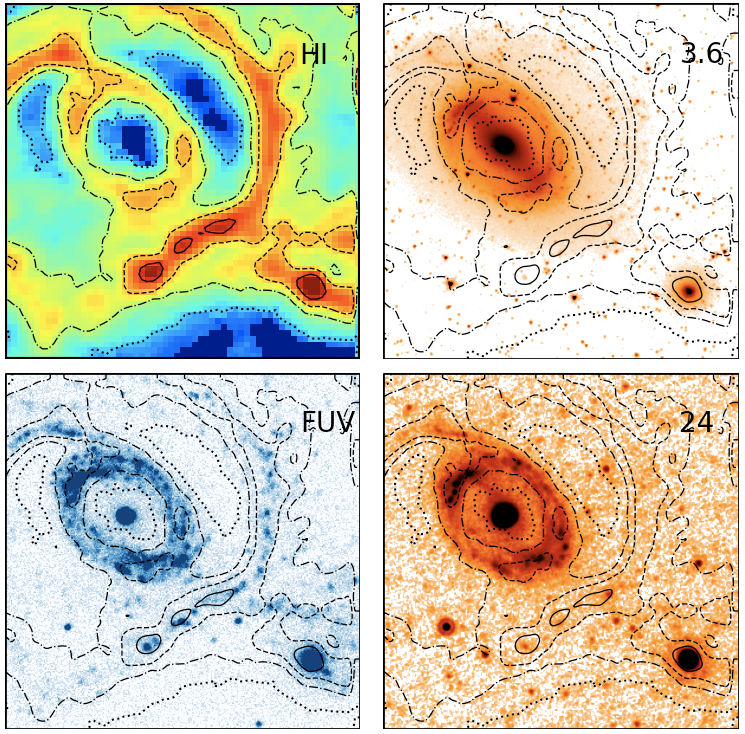}
    \caption{A multi-wavelength view of the inner portion of the NGC~1512/1510 system.  Top left: Our new \hi\ total intensity map shown with a linear colour stretch and with \hi\ mass surface density levels of 1, 3, 5, 8~\msun~pc$^{-2}$ represented by the dotted, dash-dot, dashed and solid lines, respectively.  These contours are shown in all panels.  Top right: Spitzer 3.6~$\mu$m image.  Bottom left: GALEX far-ultraviolet image.  Bottom right: Spitzer 24~$\mu$m image.  There exist clear correlations between the \hi\ flux seen in our new high-resolution MeerKAT map and the starlight observed at various wavelengths.} 
    \label{inner_HI}
\end{figure*}

The unique combination of high spatial resolution and good surface brightness sensitivity offered by our new imaging provides us with the clearest views of the system's central \hi\ content.  Here, we compare the \hi\ and stellar content of the central parts of the NGC~1512/1510 system.  The top left panel of Figure~\ref{inner_HI} shows the inner part of our \hi\ total intensity map with  \hi\ mass surface density  levels of 1, 3, 5, 8~\msun~pc$^{-2}$ represented by the dotted, dash-dot, dashed and solid lines, respectively.   The top right panel shows the Spitzer 3.6~$\mu$m view of the evolved stellar populations of NGC~1512 and NGC~1510.   Our new \hi\ imaging clearly shows the entire stellar bulge of NGC~1512 to be contained within a central \hi\ depression consisting of surface densities below 1~\msun~pc$^{-2}$, spanning a total area of $\sim11.6$~kpc$^2$.  The \hi\ has presumably been depleted by high levels of star formation.  However, in stark contrast in this regard is NGC~1510 - its stellar component is clearly co-located with an \hi\ over-density of very high mass surface densities.  In fact, its \hi\ surface densities of $\sim 12.5$~\msun~pc$^{-2}$ are the highest seen in our entire map. The enhanced \hi\ content of this region is surely linked to the star-bursting nature of NGC~1510.  

The bottom left panel of Figure~\ref{inner_HI} shows the GALEX far-ultraviolet (FUV) view of the region.  While optical images of NGC~1512/1510 typically detect only the starburst ring of NGC~1512, and NGC~1510, the FUV image reveals a much more complex morphology related to the interaction history of the two galaxies.  In the FUV image, the starburst ring of NGC~1512 is very prominent.  Comparing it to the \hi\ map, it is clear that the entire ring is associated  with an \hi\ annulus that is delimited by a surface density of $\sim 3$~\msun~pc$^{-2}$.  The semi-major axis of this elliptical \hi\ feature is $\sim 165''$ (9.4~kpc) and the total \hi mass contained with it $\sim 2.4\times 10^8$~\msun.  The regions of the starburst ring with the highest FUV fluxes are associated with higher \hi\ surface densities.  A prominent arm-like feature is located in the region between the centres of NGC~1512 and NGC~1510.  In the FUV map it is seen as a string of bright  clumps spanning an azimuthal angle range of nearly 180 degrees.  It is clear that this FUV arm is closely associated with a corresponding \hi\ arm with mass surface densities greater than 5~\msun~pc$^{-2}$.  Furthermore, the cores of the FUV clumps are  associated with clumps of higher \hi\ mass surface densities.  The correspondence between FUV and \hi\ flux in the NGC~1512/1510 system is therefore notable, and is investigated in more detail in Section~\ref{SF_laws}. 

We use Spitzer 24~$\mu$m imaging to trace the dust-obscured star formation rates of the galaxies.  In the bottom right panel of Figure~\ref{inner_HI}, we show the 24~$\mu$m image of the NGC~1512/1510 system.  The nuclear ring of NGC~1512 is expected to contain a significant amount of dust, and is indeed very bright in the 24~$\mu$m image.  However, its spatial extent is very similar to that seen in the FUV image, and the entire ring seen at 24~$\mu$m is again contained within the \hi\ flux annulus that is delimited by a surface density of 3~\msun~pc$^{-2}$.  

An alternative explanation for the presence of the central \hi\ depression surrounded by the \hi\ annulus is that of an inner Linblad resonance associated with the stellar bar of NGC~1512.  Very clear from the top right panel in Figure~\ref{inner_HI} is the fact that the \hi\ annulus is located at the end of the stellar bar. It may be the case that the \hi\ from the very inner portions of the galaxy has been concentrated into orbits that correspond to an inner Linblad resonance at the end of the bar.  An inner ring that encircles a bar and falls inside spiral arms is often identified with inner 4:1 Linblad resonance.  However, the low spectral resolution of our imaging prevents an investigations into such a possibility. 

\subsection{Measurements of extended  \hi\ features}
Having discussed the distribution of \hi\ in the vicinity of the stellar components of NGC~1512 and NGC~1510, we use this section to present some detailed measurements of several \hi\ features in the outer disk.  

Numerical simulations serve as a useful means of understanding the processes driving galaxy interactions.  Several investigators have shown that tidal perturbations of gas-rich disk galaxies result in rapid gas inflows, which occurl as a results of a rapid transfer of angular momentum from the gas to the stars.  Gas that is able to retain its angular momentum typically collects in extended disks which can consist of extended tails.  Dynamical modelling of galaxy mergers is typically carried out by finding a simulation that closely matches the observed morphology and kinematics of an interacting system (e.g., \citealt{Holincheck_2016}). Large-scale features as traced by \hi\ observations most typically used to reconstruct the orbit trajectories and disturbed morphologies of pairs of interacting galaxies.  The various length and mass measurements we present in this section can be used to constrain parameters of numerical simulations aimed at modelling the interaction history of the NGC1512/1510 system.

\begin{figure*}
\centering
	\includegraphics[width=1.7\columnwidth]{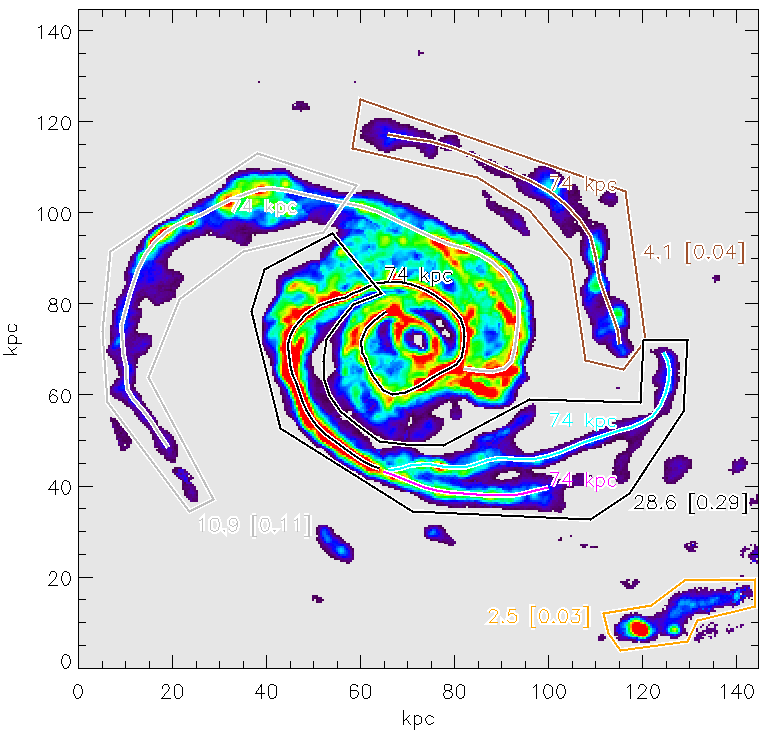}
    \caption{\hi\ total intensity map of NGC~1512/1510 shown using a linear colour stretch.  Curves coloured black, grey, magenta, cyan and brown have been visually overlaid in order to roughly trace the main \hi\ arms of the system.  The lengths of the arms are indicated in units of kpc and are also presented in the top part of Table~\ref{arm_table}.  Polygons coloured black, grey and brown have been visually overlaid to delimit the portions of the arms outside of the inner \hi\ disk.  The orange polygon delimits the \hi\ emission of the tidal dwarf galaxy.  The \hi\ masses of the polygon-delimited regions are shown in units of 10$^8$~\msun.  Also shown (in square brackets) are the masses of the polygon-delimited relative to the system's total \hi\ mass. These quantities are also shown  in the bottom part of Table~\ref{arm_table}.} 
    \label{HI_segments}
\end{figure*}

Figure~\ref{HI_segments} shows a saturated version of our \hi\ total intensity map - the colour scale spans the mass surface density range 0 to 5~\msun~pc$^{-2}$.  In this map, the high-density ridges of the system's prominent spiral arms are more clearly seen, which allows us to visually trace the path of each arm in order to roughly measure its length.  The arms all originate from, or close to, the nuclear ring of NGC~1512.  The first arm we consider is the one originating in the region between the optical components of NGC~1512 and NGC~1510.  This arm initially extends west and then wraps around anti-clockwise until it breaks out of the main \hi\ disk.  The path we visually trace for it is the grey curve in Figure~\ref{HI_segments}, and is measured to be 161~kpc in length.  Our new MeerKAT imaging traces this arm over a larger azimuthal angle  than existing image sets do.  The most prominent spiral arm of the system emerges from the north east portion of the inner \hi\ annulus seen in NGC~1512, and has its path coloured black in Figure~\ref{HI_segments}.  This arm is tightly wrapped through $\sim 360$ degrees within the main \hi\ disk but then breaks out to form the large southern spiral arm.   The length of the black path we visually trace is 129 kpc.  At the end of the black path (almost directly south of the stellar bulge of NGC~1512), this prominent \hi\ arm bifurcates.  This is clearly seen in our high-resolution MeerKAT maps, but not in existing \hi\ maps. The two bifurcations are represented by the cyan and magenta curves in Figure~\ref{HI_segments}, and have lengths of 73 kpc and 36 kpc, respectively.  The arm then extends to the north west of the galaxies.  The path  of this remaining portion is represented by the brown curve, and is measured to be  75~kpc in length.  The entire arm has a path length of approximately 129~kpc~+~73~kpc~+75~kpc~=~277~kpc.  This makes it an extremely extended feature that must be as a result of the gravitational interaction between NGC~1512 and NGC~1510.  The lengths of the various arms we trace are presented in the top half of Table~\ref{arm_table}.  

\begin{table}
	\centering
	\caption{Measurements of \hi\ arms and regions.}
	\label{arm_table}
	\begin{tabular}{ccccc} 
		\hline
		feature		& 	colour	&	length	& 	mass		&  	fractional mass\\
					&			&	[kpc]		&	[10$^8$~\msun]	&				\\
		\hline
		arm			&	grey		&	161		&	--			&	--			\\
		arm			& 	black		&	129		&	--			&	--			\\
		arm			&	brown	&	75		&	--			&	--			\\
		arm			&	magenta	&	36		&	--			&	--			\\
		arm			&	cyan		&	73		&	--			&	--			\\
		polygon		&	grey		&	--		& 	10.9			&	0.11			\\
		polygon		&	black		&	--		& 	28.6			&	0.29			\\
		polygon		&	brown	&	--		& 	4.1			&	0.04			\\
		polygon		&	orange	&	--		& 	2.5			&	0.03			\\
		\hline
	\end{tabular}
\end{table}

In order to contextualise our \hi\ arm length measurements, we compare our measurements to those made by \citet{honig_2015} for four nearby, nearly face-on, late-type spiral galaxies. Those authors measured the positions of many H\,{\sc ii} regions within the galaxies to trace their arms, and then fit log-periodic spiral models to segments of each arm.  For each spiral arm within a galaxy, they present in their Tables 2 to 5 a set of azimuthal ranges and  corresponding mean radius measurements.  The two prominent and symmetrical arms of NGC~3184 have lengths of $\sim 12.5$~kpc and $\sim 9.5$~kpc.  The southern arm in NGC~628 extends $\sim 35$~kpc.  NGC~5194 is undergoing an interaction with NGC~5195.  The shorter of its two prominent spiral arms spans $\sim 20$~kpc while the longer arm that extends from the bulge of NGC~5194 all the way to NGC~5195 is $\sim 40$~kpc in length.  Admittedly, the study of \citet{honig_2015} is based on H$\alpha$ and $B$-band imaging.  The spiral features they measure are not expected to be as extended as they might be if such a study were carried out using \hi\ imaging.  Nevertheless, it clear that the spiral arms we measure in the \hi\ distribution of NGC~1512/1510 are particularly extended.

\begin{figure}
\centering
	\includegraphics[width=\columnwidth]{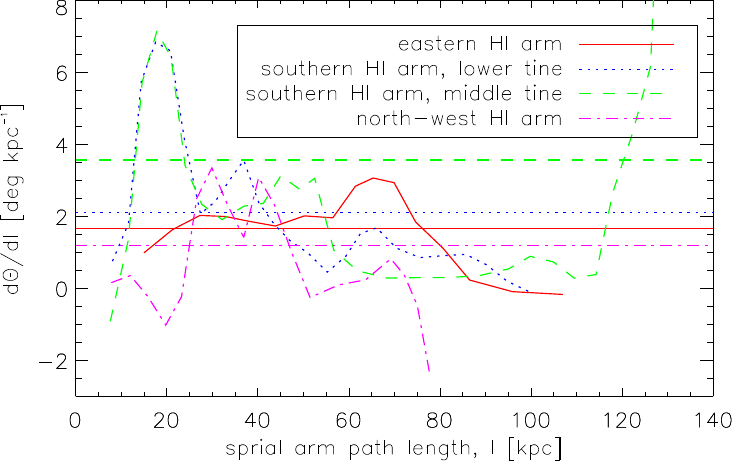}
    \caption{Rate of change of pitch angle with respect to path length for the main \hi\ spiral arms of the NGC~1512/1510 system.}
    \label{dtheta}
\end{figure}

We used our visually-estimated  lengths of the main spiral arms in the NGC~1512/1510 system to calculate the rate at which the spiral pitch angle of each arm is changing as a function of path length.   These measurements are presented in Figure~\ref{dtheta}.  The average rate at which the absolute value of the pitch angle changes with path length $\left({d\theta\over dl}\right)$ is less than $\sim 3.5^{\circ}$~kpc$^{-1}$ in all of the main arms.  Consider the portion of the southern arm delimited by the black polygon in Figure~\ref{HI_segments}, the highest values of ${d\theta\over dl}$ occur within the first $\sim 30$~kpc,  peaking at a value of $\sim 7^{\circ}$~kpc$^{-1}$.  The portion of the eastern arm delimited by the grey polygon in Figure~\ref{HI_segments} has ${d\theta\over dl}\approx 2^{\circ}$~kpc$^{-1}$ over the first $\sim 60$~kpc, after which it slightly rises and then decreases to zero.  The north-west arm (dlimited by the brown polygon in Figure~\ref{HI_segments}) has ${d\theta\over dl}$ close to zero over the first and last thirds of its path length, while ${d\theta\over dl}\approx 2^{\circ}$~kpc$^{-1}$ over the central third of its path length.

We also measure how much H\,{\sc i} mass is contained within several visually-delimited portions of the NGC~1512/1510 system.  These regions are also shown in Figure~\ref{HI_segments}, and their associated masses are summarised in lower half of Table~\ref{arm_table}.  We first consider the eastern \hi\ arm from the point at which it emerges from the main \hi\ disk of NGC~1512.  This region is  delimited by a grey polygon in Figure~\ref{HI_segments} and has an \hi\ mass of $~10.9\times 10^8$~\msun, which amounts to $\sim11$~per~cent of the \hi\ mass of the entire system.  Next, we consider the southern \hi\ arm, also from the point at which it emerges from the central \hi\ disk, all the way to the end of its bifurcated portion extending to the western half of the image.  This region, delimited by a black polygon in Figure~\ref{HI_segments}, contains $~28.6\times 10^8$~\msun\ of \hi, which is nearly 30~per~cent of the system's total \hi\ mass.  The north west extension of this \hi\ arm is the third region we consider, delimited by a brown polygon in Figure~\ref{HI_segments}.  It has an \hi\ mass of $~4.1\times 10^8$~\msun.  Overall, roughly equal amounts of \hi\ are contained within the system's main \hi\ disk and its spiral arms.  The fact that so much mass is contained within the tidally-induced \hi\ arms indicates the system to be in an advanced state of interaction.  

Finally, we consider the cloud seen in the bottom right corner of our total intensity map, delimited by the orange polygon.  This is likely a tidal dwarf galaxy in the process of forming.  It has an \hi\ mass of $~2.5\times 10^8$~\msun, which is nearly 3~per~cent of the system's \hi\ mass.  Its peak mass surface densities of $\sim 8$~\msun~pc$^{-2}$  are as high as those seen in the dense ridges of the system's prominent spiral arms.  \citet{koribalski_2009} use GALEX FUV and NUV images to demonstrate the presence of star formation in this tidal dwarf galaxy candidate. They estimate the FUV -- NUV colours of the cloud in order to derive an average age of at least 150 Myr for the young stellar population.

\section {\hi\ and FUV correlation}\label{SF_laws}
\begin{figure*}
\centering
	\includegraphics[width=1.7\columnwidth]{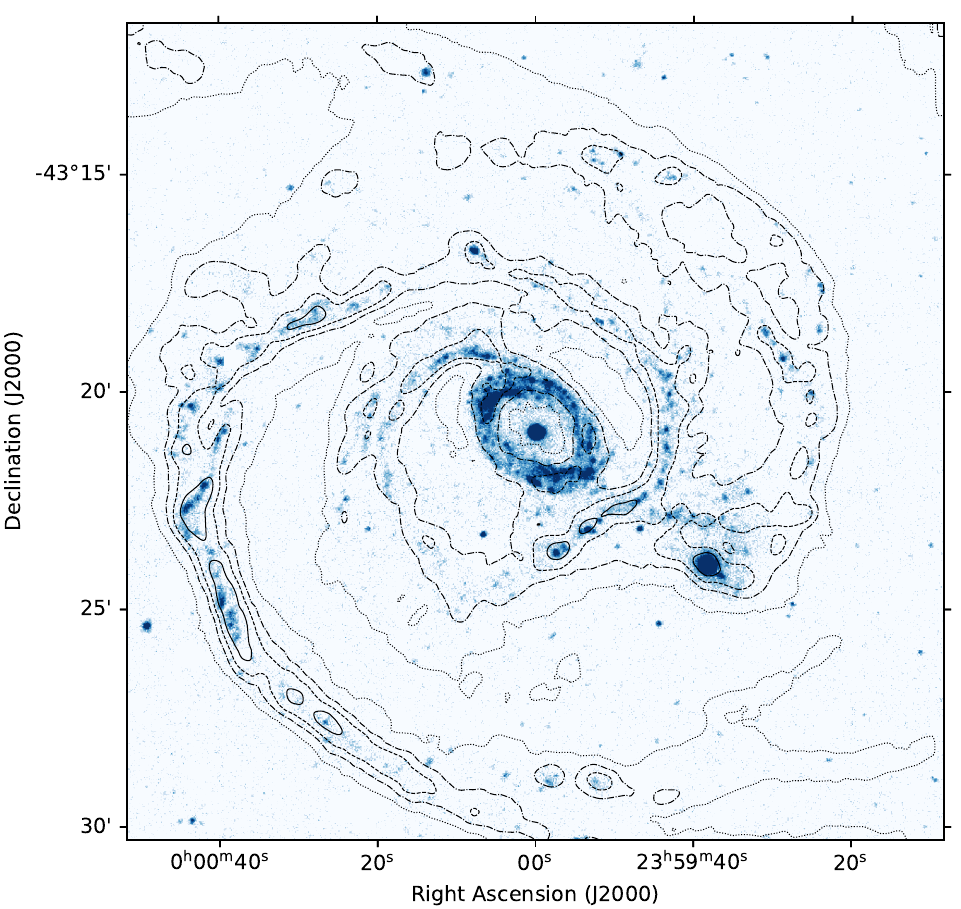}
    \caption{GALEX far-ultraviolet image showing the extended star-forming disk of the NGC~1512/1510 system.  The FUV arms extend up to $\sim 30$~kpc from the optical centre of NGX~1512.  Overlaid in black are \hi\ contours at the same levels as those shown in Fig.~\ref{mom0_panel} and Fig.~\ref{inner_HI}.  Their exists a clear correlation between \hi\ and FUV flux in the system.  We quantify the correlation and show the results in Fig.~\ref{fflux}.}
    \label{XUV}
\end{figure*}

While optical and infrared images of the system offer no obvious evidence of an interaction history between the two galaxies, the FUV image from GALEX certainly does.  Figure~\ref{XUV} presents the full GALEX image of NGC~1512/1510.  The system has a large FUV disk which,  relative to the optical centre of NGC~1512, extends out to a radius of $\sim 30$~kpc.  Very noticeable is the spatial correlation between \hi\ and FUV emission over a large range of scales. 

Rather than use only the GALEX FUV map to trace the star formation, we include the Spitzer 24~$\mu$m map as a measure of the dust-obscured star formation rate, and combine it with the the FUV map according to the prescription offered as Equation D11 in \citet{THINGS_leroy} in order to produce a map of the total star formation in the system.  Before combining the maps, we smoothed each one to the spatial resolution of our \hi\ map, and re-gridded each of them to have the same 7.5~arcsec pixel scale. Our final total star formation rate map can therefore be directly compared to our \hi\ map.  All of the system's 24~$\mu$m emission is actually contained within the starburst ring of NGC~1512, and within NGC~1510 (as shown in Figure~\ref{inner_HI}, bottom right panel).  Therefore, the outer parts of our total star formation rate map reduce to the FUV map.  In order to clearly display the complex distribution of star formation flux in the extended FUV arms of the NGC~1512/1510 system, we choose to show the GALEX FUV image in Figure~\ref{XUV} instead of the total SFR map.  However, below, we do compare the  total SFR map to our \hi\ total intensity map. 

\begin{figure}
\centering
	\includegraphics[width=1\columnwidth]{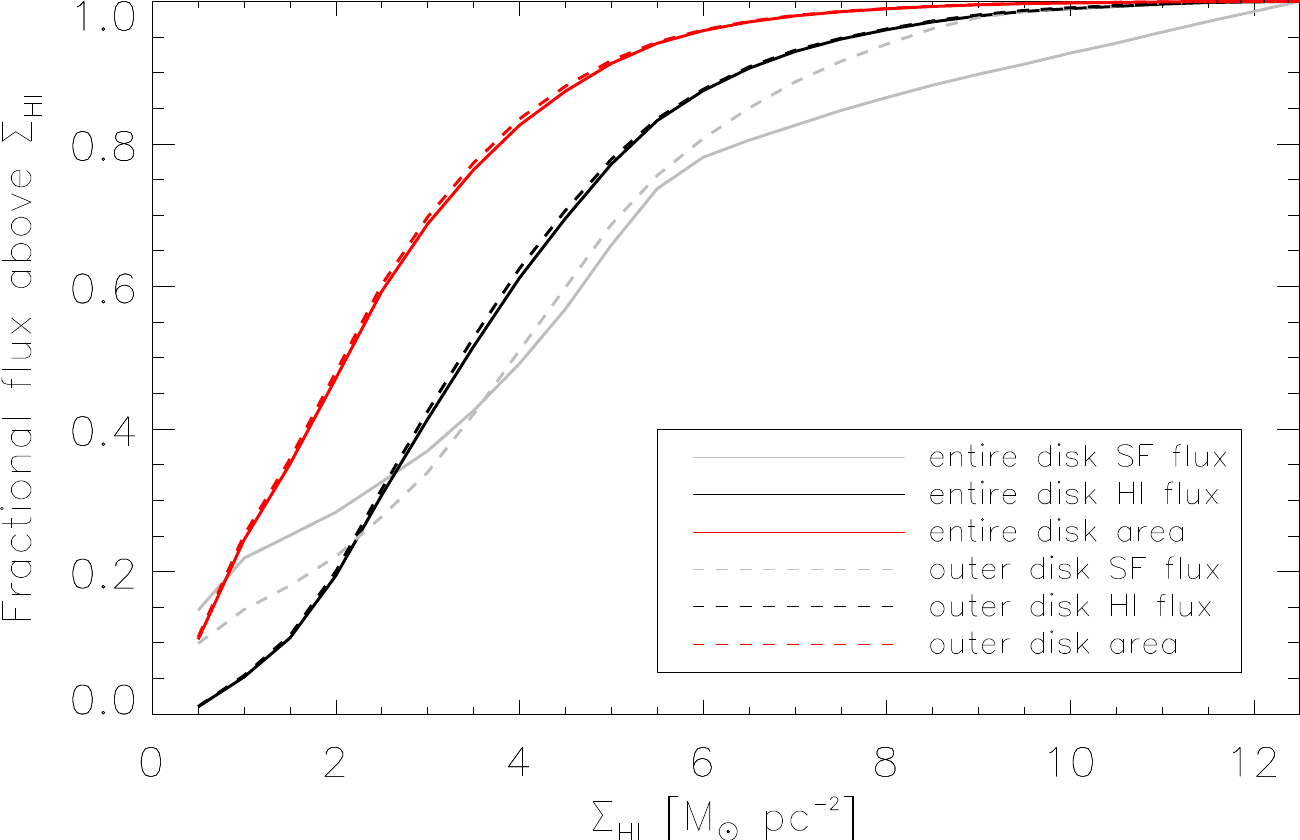}
    \caption{Distribution of fractional FUV and \hi\ flux as a functions of \hi\ mass surface density for the entire disk (solid curves) and outer disk (dashed lines) of NGC~1512.  The black/grey curves represent the fraction of \hi/FUV flux below a given surface density limit, while the red curves represent the corresponding fractional area of the \hi\ distribution.  For the outer disk of NGC~1512 there exists a notable correlation between star formation and \hi\ content.}
    \label{fflux}
\end{figure}

We measure the fractional amounts of total star formation rate and \hi\ flux corresponding to \hi\ mass surface densities ($\Sigma_\mathrm{HI}$) below a particular limit, and then compare the quantities to each other.  We use $\Sigma_\mathrm{HI}$  limits of 0.0, 0.5, 1.0, ..., 12.0, 12.5~\msun~pc$^{-2}$.  This covers the full dynamic range of our \hi\ map.  We repeat the experiment twice: once by considering all spatial pixels within the maps, and again by excluding the regions close to the optical components of the galaxies.  Figure~\ref{fflux} shows the results.  The black solid curve represents the fractional amount of \hi\ flux above a given $\Sigma_\mathrm{HI}$.  The large majority of the system's \hi\ mass is associated with low $\Sigma_\mathrm{HI}$ values.  Approximately 40~per~cent is observed at $\Sigma_\mathrm{HI}\lesssim 3$~\msun.  In Figure~\ref{mom0_panel}, we see this surface density to generally mark the edges of the high-intensity ridges of the spiral arms.  The red solid curve in Figure~\ref{fflux} is the cumulative fractional mass curve for a uniform flux distribution. In order to create it, we measured the fractional area of our \hi\ map below a given $\Sigma_\mathrm{HI}$ value. 

The grey solid curve in Figure~\ref{fflux} represents the cumulative fractional flux in our total star formation rate map.   Very clear is the fact that over the range $\Sigma_\mathrm{HI}\sim 2$~\msun~pc$^{-2}$ to $\sim 6$~\msun~pc$^{-2}$ the grey solid curve traces the black solid curve much closer than it does the red solid curve.  In other words, stellar flux is not uniformly distributed, it is tracing the \hi\ distribution.  However, at surface densities above $\sim 6$~\msun~pc$^{-2}$, the grey solid curve begins to significantly depart from the black solid curve.  This is due the extreme nature of the starburst ring in NGC~1512.  Flux from the ring constitutes a large fraction of the system's total star formation rate. However, the system's highest \hi\ mass surface densities are not correspondingly located in the starburst ring.  Rather, as can be seen in Figure~\ref{mom0_panel}, the highest \hi\ flux densities occur at the location of NGC~1510 , the region between NGC~1512 and ~NGC~1510, as well as the large \hi\ arm that extends to the south of NGC~1512.  Within the starburst ring region of NGC~1512 there is presumably a high content of molecular hydrogen that is playing the dominant role in regulating the star formation activity.  

In order to remove from our analysis the complexities associated with the nuclear ring, we mask it together with the region centred on NGC~1510.  We repeat our cumulative flux measurements and present the results as dashed curves in Figure~\ref{fflux}.  For the extended FUV disk of the system, we see even stronger evidence of a correlation between star formation and \hi\ mass. The grey dashed and black dashed curves trace one another closely over a larger range of \hi\ mass surface densities. This is quantitative confirmation of the spatial correlation that is obvious from our comparison of the FUV and \hi\ maps shown in Figure~\ref{XUV}.  Thus, for the extended star-forming disk of the NGC~1512/1510 system, it seems that high-mass star formation occurs where \hi\ is located.  In spite of the complex \hi\ morphology of the NGC~1512/1510 system, the presence of \hi\ is clearly a precondition for star formation.  Similar results are found by \citet{bigiel_2010} for M83.  They report that their ``FUV and \hi\ maps show a stunning spatial correlation out to almost 4 optical radii''.  They show the \hi\ depletion time in the outer disk to be about 100~Gyr, and comment that it is likely the inefficient build up of molecular clouds that is the bottleneck for forming stars at large radii. 



\section {Summary}\label{sec_summary}
We have presented MeerKAT \hi-line imaging of the nearby interacting galaxy pair NGC~1512/1510.  The powerful combination of high spatial resolution ($26.8''\times 20.1''$) and flux sensitivity offered by MeerKAT data provides us with a new view of the \hi\ content of the system, especially its distribution and its links to star formation activity.

Our primary data set is a new \hi\ total intensity map in which we see the entire \hi\ morphology to be dominated by the tidally-induced \hi\ arms. Our map yields a total \hi\ mass $M_\mathrm{HI} = 9.74^{11.7}_{8.01}$~\msun.  For the first time, we are able to resolve the spatial distribution of \hi\ mass near the centre of the system.  The starburst ring seen in optical images of NGC~1512 is clearly associated with a corresponding \hi\ annulus that is delimited by a mass surface density of $\sim 3$~\msun~pc$^{-2}$, while the bright stellar bulge of NGC~1512 is co-located within a prominent central \hi\ depression where \hi\ mass is  concentrated less than 1~\msun~pc$^{-2}$.  NGC~1510, however, has its bright star-bursting core co-located with the highest \hi\ surface densities we detect ($\sim 12.5$~\msun~pc$^{-2}$).  
 
 The improved spatial resolution of our new \hi\ total intensity map allows us quantify the correlation between \hi\ mass and emission from high-mass stars on $\sim 1.5$~kpc length scales.  When we exclude the extreme regions close to the optical components of the two galaxies, we find a  clear correlation between \hi\ and far-ultraviolet flux, similar to what is seen in other nearby galaxies.  Hence, in spite of the  history of interaction between NGC~1512 and NGC~1510, the \hi\ within the system seems to be playing an important role in setting the pre-conditions required for high-mass star formation.  
 
 The maps we generate and show in this work demonstrate the raw imaging power of MeerKAT, and exemplify the studies that will soon be carried out routinely as parts of large nearby galaxy surveys, and at higher redshifts.  

\section{Acknowledgments}
The MeerKAT telescope is operated by the South African Radio Astronomy Observatory (SARAO), which is a facility of the National Research Foundation, an agency of the Department of Science and Innovation. EE's research is supported by SARAO.  The authors acknowledge the use of computing facilities of the Inter-University Institute for Data Intensive Astronomy (IDIA) for part of this work. MG acknowledges support from IDIA and was partially supported by the Australian Government through the Australian Research Council's Discovery Projects funding scheme (DP210102103.  RPD acknowledges the South African Research Chairs Initiative of the Department of Science and Innovation and the National Research Foundation.  

\section{Data availability}
Requests for access to the data products presented in this work may be submitted via email to the corresponding author.



\bsp	
\label{lastpage}
\end{document}